\documentclass[11pt]{article}
\usepackage{graphicx}
\usepackage{amssymb}
\usepackage{amscd}
\usepackage{mathrsfs}
\usepackage{longtable,lscape}
\usepackage{amsthm}
\usepackage{amsfonts}
\usepackage{amsmath}
\usepackage{bbm}
\usepackage{float}
\usepackage{url}
\usepackage{hyperref}
\usepackage[margin=0.5cm,font=footnotesize]{caption}
\usepackage{lineno}
\usepackage[round]{natbib}
\usepackage{appendix}
\usepackage{subfig}

\begin{document}
\title{\bf Entanglement as a Method to Reduce Uncertainty}
\author{Diederik Aerts, Jonito Aerts Arg\"uelles, Lester Beltran, Suzette Geriente\footnote{Center Leo Apostel for Interdisciplinary Studies, 
        Free University of Brussels (VUB), Krijgskundestraat 33,
         1160 Brussels, Belgium; email addresses: diraerts@vub.be,jonitoarguelles@gmail.com,lestercc21@yahoo.com,sgeriente83@yahoo.com} 
        $\,$ and $\,$  Sandro Sozzo\footnote{Department of Humanities and Cultural Heritage (DIUM) and Centre CQSCS, University of Udine, Vicolo Florio 2/b, 33100 Udine, Italy; email address: sandro.sozzo@uniud.it}              }

\date{}

\maketitle
\begin{abstract}
\noindent
In physics, entanglement `reduces' the entropy of an entity, because the (von Neumann) entropy of, e.g., a composite bipartite entity in a pure entangled state is systematically lower than the entropy of the component sub-entities. We show here that this `genuinely non-classical reduction of entropy as a result of composition' also holds whenever two concepts combine in human cognition and, more generally, it is valid in human culture. We exploit these results and make a `new hypothesis' on the nature of entanglement, namely, the production of entanglement in the preparation of a composite entity can be seen as a `dynamical process of collaboration between its sub-entities to reduce uncertainty', because the composite entity is in a pure state while its sub-entities are in a non-pure, or density, state, as a result of the preparation. 
We identify within the nature of this entanglement a mechanism of contextual updating and illustrate the mechanism in the example we analyze. 
Our hypothesis naturally explains the `non-classical nature' of some quantum logical connectives, as due to Bell-type correlations.
\end{abstract}

\section{Introduction\label{intro}}
There is increasing agreement in the scientific community to look at quantum entanglement as a `resource' rather than a `conundrum', entanglement and its ensuing nonlocality and nonseparability being the distinctive aspects of quantum theory \citep{wootters1998,prevedelatal2007,zhangetal2016,trapaniparis2017}.\footnote{In this regard, it is worth noticing that the Nobel Prize in Physics 2022 was awarded to Alain Aspect, John Clauser and Anton Zeilinger ``for experiments with entangled photons, establishing the violation of Bell inequalities and pioneering quantum information science'', see the website \url{https://www.nobelprize.org/prizes/physics/}.} This change of perspective is partially due to the advent of quantum technologies, where entanglement is used to, e.g., encrypt, encode, communicate, and teleport, thus implementing computational and informational tasks \citep{nielsenchuang2000,bub2007,timpson2013}. However, we believe that this is not the only reason and there is a more fundamental one, which can be traced back to the foundations of quantum theory. The scope of this paper is exactly to demonstrate that entanglement can be used as a tool to reduce uncertainty.

In the last two decades, it has more and more become evident that entanglement is not peculiar of micro-physical systems, or entities,\footnote{We use the term `entity' as a synonym of `system' across the paper.} as electrons and photons \citep{epr1935,bohm1951,bell1964,chsh1969,brunner2014}, but it can also be brought out in macroscopic entities as vessels of water \citep{aerts2010}, in combined concepts as \emph{The Animal Acts} \citep{aertssozzo2011,aertssozzo2014,beltrangeriente2018,aertsetal2019,arguellessozzo2020,aertsetal2021a,aertsetal2023a}, in human culture \citep{aertsetal2023b} and, more generally, whenever any two (or more) entities can be `connected' in a way that is similar to how two spin-1/2 quantum entities are connected when they are prepared in an entangled state.\footnote{The fact that such entities are non-separated is crucial to be able to represent them in Hilbert space, as this representation fails when the entities are instead `separated' \citep{aerts1982}.}

As a matter of fact, a significant violation of Bell's inequalities \citep{bell1964,chsh1969} has been empirically observed in a variety of domains other than micro-physics (see, e.g., \citet{aspectdalibardroger1982,vienna2013,urbana2013}), from macroscopic physics (see, e.g., \citet{aerts2010}) to the combination of concepts in both cognitive tests on human participants \citep{aertssozzo2011,aertssozzo2014,aertsetal2023a} and information retrieval tests on the World Wide Web (document retrieval \citep{beltrangeriente2018,aertsetal2021a} and image retrieval \citep{arguellessozzo2020}).\footnote{The research programme that applies quantum structures to model phenomena outside of micro-physical domains is known as `quantum interaction' and has already reached important milestones (see, e.g., \citet{aerts2009a,busemeyerbruza2012,aertsbroekaertgaborasozzo2013,aertsgaborasozzo2013,havenkhrennikov2013,dallachiaragiuntininegri2015a,dallachiaragiuntininegri2015b,
melucci2015,aertssozzo2016,aertssozzoveloz2016,blutnerbeimgraben2016,aertsetal2018,pisanosozzo2020,aertsetal2021b}).} In these tests, we have in particular investigated the conceptual combination \emph{The Animal Acts}, considered as a composite entity made up of the conceptual sub-entities \emph{Animal} and \emph{Acts}, where the word `acts' refers to the action of the animal that emits a sound. In this case, the sub-entities \emph{Animal} and \emph{Acts} concur to entangle when they combine, due to the emergent meaning carried by the combined entity \emph{The Animal Acts}, which determines non-classical statistical correlations when experiments are performed, exactly as it occurs with micro-physical entities as electrons and photons. 

Recently, we have found an important relationship between entanglement and entropy, which holds in both physics and cognition realms \citep{aertsetal2023b,aertsbeltran2022}. More specifically, the von Neumann entropy of the composed entity, be it the composition of two spin-1/2 quantum entities or the combined conceptual entity \emph{The Animal Acts}, is always lower than the von Neumann entropy of the component sub-entities, be them the two spin-1/2 quantum entities or the conceptual entities \emph{Animal} and \emph{Acts}. This reduction of entropy is due to the fact that the composed entity is in a pure entangled state while the sub-entities are in a non-pure, or density, state. Equivalently, independently of the realm, micro-physical or cognitive, the preparation of a composed entity in a pure entangled state reduces the `uncertainty' associated with the component sub-entities -- if the composite entity is in the singlet spin state, the uncertainty of its component parts is maximal. This result allows us to put forward a new way of looking at entanglement as a `dynamical process of uncertainty reduction in which the sub-entities actively collaborate to produce a pure state of the whole', and fundamentally differs from the original conception of entanglement as an `almost instantaneous spooky action at a distance produced by far away performed experiments' (see, e.g., \citet{epr1935}). 

Interestingly enough, the idea of entanglement as a collaborative process finds applications in human culture too. For example, the process of `hunting' can be seen as a dynamical process of entanglement, in which two (or more) hunters have to actively collaborate to produce an overall entangled state in such a way to reduce the uncertainty of the whole hunting situation and reach their common objective \citep{aertsetal2023b}.  

Finally, the presence of entanglement has also an impact on 
some of the connectives of quantum logic, which can again be explained in terms of the new way of looking at entanglement that we put forward in this paper. Indeed, due to the Bell-type correlations created whenever a composed entity is prepared in the singlet spin state, some of the quantum logical connectives, in particular, the connectives corresponding to the `inclusive or', `exclusive or' and `bi-conditional' of classical logic, behave in a non-classical way \citep{aerts1982,aertsdhondtgabora2000}, in the sense that the proposition obtained through these connectives can be true without any of the component propositions being either true or false. Our explanation for this is that the above-mentioned reduction of uncertainty produced by a pure entangled state also makes it possible the composite proposition to have a `more classical logical structure' than the component propositions.

For the sake of completeness, we summarise the content of this paper in the following.

In Sect. \ref{physics}, we present the way in which quantum physicists have historically revealed entanglement by means of a violation of Bell's inequalities in a Bell-type test. Then, we illustrate the relationship between entanglement and von Neumann entropy as presented in modern manuals of quantum physics and quantum information, and formulate our hypothesis of entanglement as an active process of uncertainty reduction as an alternative explanation to the spooky action at a distance hypothesis.

In Sect. \ref{cognition}, we review our theoretical and empirical analysis of entanglement in the conceptual combination \emph{The Animal Acts} and show that entanglement and von Neumann entropy exhibit the same relationship that we have discussed in Sect. \ref{physics}, namely, entropy decreases as a result of composition. This allows us to consider the implications of our main hypothesis with respect to meaning and natural language. Meaning cannot be carried by product states but, rather, by entangled states.

In Sect. \ref{culture}, we present the `hunting example' as a paradigmatic situation in human culture where the collaboration between individuals to reach a common objective can be structurally seen as an entanglement process, and indicate that this way of looking at collaborations could be extended to other domains of human culture.  

In Sect. \ref{quantumlogic}, we finally illustrate the implications of our main hypothesis at the level of logic, and prove that the non-classical nature of quantum logical connectives is a fundamental consequence of the Bell-type statistical correlations exhibited by composite entities in entangled states.

\section{Entanglement and entropy in micro-physics \label{physics}}
We review in this section the relationship between entanglement and von Neumann entropy that holds in quantum physics and quantum information theory \citep{nielsenchuang2000,bub2007,timpson2013}. Making this relationship explicit will allow us to put forward a new and unitary hypothesis on the nature of quantum entanglement in physics, cognition and culture (see also Sects. \ref{cognition} and \ref{culture}).

Let us preliminarily consider what physicists usually call a `Bell--type test' \citep{bell1964,chsh1969}. Let $S_{12}$ be a composite bipartite\footnote{For the sake of simplicity, we limit ourselves to consider bipartite entities in pure states and apply the notions of entanglement and von Neumann entropy to such composite entities. The discussion of the more general case of multipartite mixed state entanglement, though mathematically 
interesting, is not relevant to the scopes of this paper.} entity and let $S_1$ and $S_2$ be two individual entities which can be recognised as parts of $S_{12}$. Then, let us consider the experiment $X$, where $X=A,A'$, with outcomes $X_i$, where $i=1,2$, that can be performed on $S_1$, and experiment $Y$, where $Y=B,B'$, with outcomes $Y_j$, where $j=1,2$, that can be performed on $S_2$, so that the coincidence experiment $XY$ can be performed on $S_{12}$ which consists in performing $X$ on $S_1$ and $Y$ on $S_2$. If the experiment outcomes $X_i$ and $Y_j$ can only be $\pm 1$, then the expected value of $XY$ becomes the correlation function $E(XY)$. It is well known that, under the reasonable, in a classical physics perspective, assumptions of locality and realism, one can prove the `Clauser-Horne-Shimony-Holt (CHSH) version' of Bell's inequalities \citep{chsh1969}, that is,
\begin{equation} \label{chsh}
-2 \le \Delta_{CHSH} \le +2
\end{equation}
where $\Delta_{CHSH}$ is the `CHSH factor' which is defined as
\begin{equation} \label{CHSH_factor}
\Delta_{CHSH}=E(A'B')+E(A'B)+E(AB')-E(AB)
\end{equation}
One can also prove that Bell's inequalities constitute a mathematical constraint to the representation of the statistical correlations that are exhibited by the parts of a composite entity by means of a classical probabilistic, i.e. Kolmogorovian, formalism (see, e.g., \citet{accardifedullo1982,pitowsky1989}). On the other side, micro-physical entities as electrons and photons generally violate inequality (\ref{chsh}). For example, let $S_1$ and $S_2$ be two spin–1/2 quantum entities and let $S_{12}$ be the composite entity up of $S_1$ and $S_2$ and prepared in the `singlet spin state' $p_{s}$. In the computational basis $\{ \vert 0\rangle=(1,0), \vert 1\rangle=(0,1) \}$ of the Hilbert space $\mathbb{C}^2$, $p_{s}$ is represented by the unit vector
\begin{equation} \label{singlet}
\vert \Psi_{s}\rangle=\frac{1}{\sqrt{2}}(\vert 01\rangle-\vert 10\rangle)
\end{equation}
where we have omitted the tensor product notation and set $\vert 01\rangle=\vert 0 \rangle \otimes \vert 1 \rangle$ and $\vert 10\rangle=\vert 1 \rangle \otimes \vert 0 \rangle$. Next, for every $X=A,B'$ and $Y=B,B'$, let $XY$ be the coincidence experiment that consists in 
measuring the spin along direction $X$ on $S_1$ and the spin along direction $Y$ on $S_2$, and suppose that the angles between directions are $(A,B)=(B,A')=(A',B')=45^{\circ}$. In this case, one easily gets that the CHSH factor in Eq. (\ref{CHSH_factor}) is equal to $2\sqrt{2}\approx 2.8284$, which is significantly greater than 2. This violation of Bell's inequalities in quantum physics has been largely confirmed empirically \citep{aspectdalibardroger1982,vienna2013,urbana2013} and reveals that micro-physical entities as electrons and photons exhibit genuinely non-classical features, that is, contextuality, nonlocality, non-separability, and what the founding fathers considered as the most important one, namely, entanglement. Entanglement, in particular, entails that micro-physical entities can be `connected' so strongly that they do not have any property, hence kind of lose their identity, within a composite entity. 

The presence of entanglement in a composite entity $S_{12}$ is particularly evident if one calculates the von Neumann entropy associated with the entities $S_1$ and $S_2$ that are parts of $S_{12}$ \citep{nielsenchuang2000}. To this end, it is more convenient to represent states in the `density operator formalism' \citep{desp1976}. In this formalism, the state of an entity is represented by a density operator $\rho$, i.e. a positive self-adjoint operator with $\textrm{Tr}\rho=1$, on the Hilbert space $\mathscr H$ associated with the entity. More precisely, a pure state is represented by a density operator $\rho=\vert\psi\rangle\langle\psi\vert$, where $\vert \psi\rangle$ is a unit vector of $\mathscr H$, hence, $\textrm{Tr}\rho^2=\textrm{Tr}\rho=1$, while a mixed state (or, proper statistical mixture) is represented by a density operator $\rho$ such that $\textrm{Tr}\rho^2<\textrm{Tr}\rho=1$.

Let us come to the von Neumann entropy, which is generally considered as a `measure of the uncertainty associated with the state of a quantum entity' \citep{nielsenchuang2000,bub2007,timpson2013}. The von Neumann entropy is defined as
\begin{equation} \label{vonnneumann}
S(\rho)=-\textrm{Tr}\rho \log_2 \rho
\end{equation}
where $\log_2$ denotes the logarithm of a number in base 2. Let us consider the spectral decomposition of $\rho$, namely, $\rho=\sum_{k=1}^{n}p_k \vert\psi_k\rangle\langle\psi_k\vert$, where, for every $k=1,\ldots,n$, $p_k\ge 0$, $\sum_{i=k}^{n}p_k=1$, and $\{\vert\psi_k\rangle \}_{k=1,\ldots,n}$ is an orthonormal basis of $\mathscr H$. Then, Eq. (\ref{vonnneumann}) becomes
\begin{equation} \label{shannon}
S(\rho)=-\sum_{k=1}^{n} p_k \log_2 p_k
\end{equation}
which coincides with the `Shannon entropy' expressing the uncertainty associated with a classical probability distribution generated by $p_1, \ldots, p_n$. 

We dedicate the rest of this section to illustrate how the von Neumann entropy relates to the entanglement situation introduced above. A general property of the von Neumann entropy is that, for every density operator $\rho$, $S(\rho) \ge 0$. In addition, $S(\rho)=0$ if and only if $\rho$ represents a pure state. Let $S_{12}$ be a composite entity in the singlet spin state $p_{s}$ represented by the density operator $\rho_{\Psi_{s}}=\vert \Psi_{s} \rangle \langle \Psi_{s} \vert$, where $\vert \Psi_{s}\rangle$ is the unit vector in Eq. (\ref{singlet}). The states of the parts $S_1$ and $S_2$ of $S_{12}$ are represented by density operators $\rho_1$ and $\rho_2$ obtained by performing the operation of partial trace of $\rho_{\Psi_{s}}$ with respect to $S_2$ and $S_1$, respectively. One then easily proves that $\rho_1=\frac{1}{2}\mathbbm{1}=\rho_2$, where $\mathbbm{1}$ is the identity operator on the Hilbert space $\mathbb{C}^2$. Hence, the von Neumann entropies of the component sub-entities $S_1$ and $S_2$ are maximal, namely, $S(\rho_1)=\log_2 2=S(\rho_2)$. In other words, whenever the composite entity is in the singlet spin state, associated with null entropy, hence null uncertainty, its parts are in a non-pure, or density, state, associated with maximal entropy, hence maximal uncertainty.\footnote{One should be careful when talking about `uncertainty' in $\rho_1$ and $\rho_2$. Indeed, while these density operators mathematically represent mixed states, the uncertainty revealed by the non-null entropy of $\rho_1$ and $\rho_2$ cannot be interpreted as `uncertainty about the state'. To express this foundationally fundamental difference, some authors use the term `improper statistical mixture' to indicate the states of parts of a composite entity in an entangled state obtained by means of partial traces \citep{desp1976}. However, it can be proved that the non-null entropy of $\rho_1$ and $\rho_2$ can be interpreted as the `uncertainty in the measurement on $S_1$ and $S_2$, respectively \citep{aertssassoli2014}. \label{uncertainty}}

The result above is absolutely general. Whenever a composite entity is in a pure entangled state and is thus associated with null von Neumann entropy, its parts are in a density state and are thus associated with positive von Neumann entropy. This property of composite quantum entities does not have a counterpart in classical physics, hence, von Neumann entropy is generally used as a `measure of entanglement' \citep{horodeckis2009}.

These considerations allow us to make a new interesting hypothesis on the nature of entanglement, namely, the production of entanglement in a preparation can be considered as a `collaborative process of reduction of uncertainty', in the sense that, in a preparation process of a pure entangled state on a composite entity, the component sub-entities, by entangling, collaborate to produce a pure state, while they are in a non-pure, or mixed, state. As a consequence, the sub-entities are affected by uncertainty, that is maximal in the case of a maximally entangled state as the singlet spin state, while the composite entity is not affected by uncertainty. If we look at entanglement in this way, the dynamics involved in the preparation process of an entangled state is the crucial aspect, while the measurements which determine a violation of Bell's inequalities only play a secondary role. 
What we mean is that by the focus placed by physicists on the question `can signals be sent by using entanglement' from Bob to Alice or vice versa, the phenomenon was too narrowly reduced to the very special case of `photons flying far away from each other yet remaining entangled', or to the question `is there spooky action at a distance or not?'. We do not mean by this that this special situation is not important, certainly not, that entanglement of spatially distant entities exists, has been experimentally verified, and signalling involved has been ruled out, undoubtedly remains a very important finding, because it reveals to us an unexpected and mysterious property of physical reality, and of space and time. But, the property of entanglement, which we examine here, is at least as important, and it consists of combining two entities into one, and such that the uncertainty on the composite entity is smaller than the uncertainty on each of the two subentities. In the extreme case, the subentities may entail total uncertainty, and their combination into the single entity may lead to total elimination of this uncertainty. And this property is not necessarily linked to whether signalling takes place or not in measurements being made, for it is of a structural nature, and directly due to the `not being a product state' of the state of the composite entity. 
The preparation of entangled photons with linear optics is insightful in this sense \citep{linearoptics2001}. This means that entanglement should not be 
only regarded 
with a focus on the question whether 
a `spooky action at-a-distance' takes place, that is, 
whether an instantaneous influence of measurements performed on far away entities exists, but, 
equally so, and perhaps even primarily so, with the focus on it being 
an `active collaboration between entities in the preparation process',
in other words, 
with a focus on
the preparations, instead of on the measurements.
What is then important is that the parts of a composite entity in some way collaborate to produce a pure entangled state, lowering uncertainty in an intrinsic and structural way, and this
will be responsible of the statistical correlations that violate Bell's inequalities. 

The new way of looking at entanglement that we have highlighted above may have a profound impact on the foundations of quantum physics. More important, it is exactly the same type of entanglement that one finds in the combination of concepts in human cognition, as we will see in the next section.

\section{Entanglement and entropy in conceptual combinations \label{cognition}}
As in Sect. \ref{physics}, we analyse in this section the relationship between entanglement and entropy that holds in the combination of two concepts, with the aim of investigating to what extent entanglement can be used as a means to reduce uncertainty. To this end, we need to preliminarily illustrate the empirical and theoretical research we have conducted in the last decade on the violation of Bell's inequalities in the conceptual combination {\it The Animal Acts}. We limit ourselves to report here the results that are needed for the scopes of this  
article. The reader who is interested in technical details can refer to Refs. \citet{aertssozzo2011,aertssozzo2014,beltrangeriente2018,aertsetal2019,arguellessozzo2020,aertsetal2021a,aertsetal2023a,aertssozzo2016,aertsetal2023b,arguelles2018} and references therein. In particular, a complete analysis on the relationship between physics and cognition in regard to entanglement can be found in Ref. \citet{aertsetal2019}.

Our research team have conducted several Bell-type tests to reveal the presence of entanglement in the combination of two concepts by means of a violation of the CHSH version of Bell's inequalities and the ensuing quantum representation in Hilbert space. More specifically, we have studied the conceptual combination {\it The Animal Acts}, which we have considered as a composite conceptual entity made up of the sub-entities {\it Animal} and {\it Acts}. As specified in Sect. \ref{intro}, `acts' refers to the sound, or noise, produced by an animal. The tests we have performed on {\it The Animal Acts} can be substantially divided into two groups: (i) cognitive tests involving human participants \citep{aertssozzo2011,aertsetal2023a}, and (ii) information retrieval tests involving search engines on the World Wide Web \citep{beltrangeriente2018,arguellessozzo2020,aertsetal2021a,arguelles2018}. Concerning (ii) the first test was a document retrieval test on the corpuses of documents  `Google Books', `Contemporary American English (COCA)' and `News on Web (NOW)' \citep{beltrangeriente2018}, while the second test was an image retrieval test using the search engine `Google Images' \citep{arguelles2018}. 

Despite the specific character of the different tests, e.g., different ways of collecting data, differences in the calculation of the relative frequencies converging to the probabilities entering Bell's inequalities, values of the CHSH factor in Eq. (\ref{chsh}), all these tests showed a common pattern, namely, all tests significantly violated Bell's inequalities. For any of those tests, the data that are relevant to the purposes of this paper are reported in Table \ref{animalactsentanglements}. In particular, the last column of the table reports the CHSH factor, where the terms in Eq. (\ref{chsh}) are calculated as follows. For every $X=A,A'$, $Y=B,B'$, 
\begin{equation}
E(XY)=\sum_{i,j=1}^{2}X_iY_jp(X_iY_j)
\end{equation}
where, for every $i=j$, $X_iY_j=+1$, for every $i \ne j$, $X_iY_j=-1$, and $p(X_iY_j)$ is the probability of obtaining the outcome $X_iY_j$ in the experiment $XY$ (see Sect. \ref{physics}).

By looking at the CHSH factor in Table \ref{animalactsentanglements}, we notice that, on the one side, some tests violate Bell's inequalities by amounts that are close to the value $2\sqrt{2}\approx 2.8284$, predicted by quantum theory and substantially confirmed in quantum physics tests (see, e.g., \citet{aspectdalibardroger1982,vienna2013,urbana2013}). But, on the other side, other tests exhibit a violation which goes even beyond $2\sqrt{2}\approx 2.8284$. The latter value is known as the `Cirel'son bound' and is typically considered as a theoretical limit to represent in Hilbert space the statistical correlations observed  in Bell-type tests by pushing all entanglement into the state of the composite entity and considering only product measurements, i.e. measurements with product eigenstates. If, on the contrary, one allows `entangled measurements', i.e. measurements where at least one eigenstate is entangled, then a Hilbert space representation is possible also for Bell-type tests violating the Cirel'son bound \citep{aertsetal2023a}. These empirical results convincingly show that the concepts {\it Animal} and {\it Acts} `entangle' when they combine to form {\it The Animal Acts}. The entanglement is due to the fact that people 
attribute meaning to the combination {\it The Animal Acts} as a whole entity, without firstly attributing meaning to {\it Animal} and {\it Acts} and then combining 
these separate meanings into a meaning for {\it The Animal Acts}. 
One way to characterize this process of meaning assignment is the following. A word carries a meaning, and a second word carries a meaning, however, the combination of these two words also carries its proper meaning, and this is not the simple combination of the two meanings of the subwords. The new emerged meaning of the combined word arises in a complex contextual way, in which the whole of the context relevant to the story plays a role. We can speak of an `updating' of contextuality each time a word is added, and this updating continues to take place until the end of the story that contains all the words.
And, this 
`contextual updating way' to 
attribute meaning has to be carried by an entangled state. 
For those familiar with the mathematical Hilbert space formalism of quantum mechanics, and the use of the tensor product for the description of compound quantum entities, they will recognize such a `contextual updating' mechanism in the mathematical procedure of describing multiple compound quantum entities. Indeed, whenever a Hilbert space of an individual quantum entity is coupled to the tensor product, new states form, as a consequence of the superposition principle, which always contain a majority of entangled states. It is these that accomplish the contextual updating in the mathematical formalism. This deep structural similarity to what takes place in human language, and what takes place mathematically in the formalism of quantum mechanics, is an additional argument for our claim that it is this structural property of entanglement that is the most important and the most significant. We believe, incidentally, that this is also the case for what entanglement represents in physics, although, as we have already noted, there the focus has been shifted to questions related to `signalling'. It would distract us too far from the topic of the present article, but we plan to return to it in future work, we think that the additional problem associated with the `spooky action' intuition in physics, rests on an incomplete analysis of the measurement process in quantum mechanics itself.
Taking into account the above, and returning to our concrete situation, we have provided a faithful quantum representation in Hilbert space for all data sets where {\it The Animal Acts} does correspond to an entangled, not a product, state.\footnote{Given a cognitive entity, e.g., a concept, a conceptual combination, or a more complex decision entity, that can undergo an empirical test in which states, contexts and non-classical probabilities of outcomes can be identified, a theoretical procedure exists to represent the entity in the Hilbert space formalism of quantum theory (see, e.g., \citet{aertssassolisozzo2016}).}

Let us now calculate the von Neumann entropy associated with the states of the composite entity {\it The Animal Acts} and its parts {\it Animal} and {\it Acts}, using the general  procedures illustrated  in Sect. \ref{physics}.

\begin{table}
\centering
\caption{We report here some of the data collected in the five Bell-type tests of entanglement on the conceptual combination {\it The Animal Acts}. For each test, we report the outcome probabilities of experiment $AB$, and the CHSH factor of the overall test.}
\label{animalactsentanglements}
\begin{tabular}{l|llll|l|l}
\hline
Tests & \multicolumn{4}{l|}{Probabilities of experiment $AB$} & Entropy & CHSH factor \\
 & $p(HG)$ & $p(HW)$ & $p(BG)$ & $p(BW)$ & $S$ &  $\Delta_{CHSH}$  \\
\hline
2011 cognitive test & $0.049$ & $0.630$ & $0.259$ & $0.062$ & $0.177$ & $2.4197$ \\
& & & & & & \\
Google Books test & $0$ & $0.6526$ & $0.3474$ & $0$  & $0.280$ & $3.41$ \\
& & & & & & \\
COCA test & $0$ & $0.8$ & $0.2$ & $0$ & $0.217$ & $2.8$ \\
& & & & & & \\
Google Images test & $0.0205$ & $0.2042$ & $0.7651$ & $0.0103$ & $0.202$ & $2.4107$ \\
& & & & & & \\
2021 cognitive test & $0.0494$ & $0.1235$ & $0.7778$ & $0.0494$ & $0.114$ & $2.79$ \\
\hline
\end{tabular}
\end{table}

Let us consider five empirical studies on {\it The Animal Acts}, as reported in the first column of Table \ref{animalactsentanglements}. We limit ourselves to illustrate how data are collected in a Bell-type test with human participants \citep{aertssozzo2011,aertsetal2023a}, for the sake of brevity.\footnote{In an information retrieval test, data are collected differently, but the experiment outcomes and the values that are computed, that is, outcome probabilities and the CHSH factor, are exactly the same as in the tests on human participants.} In a Bell-type test on human participants, a questionnaire is submitted to every participant which contains an introductory text precisely explaining entities and tasks involved in the test. Let us focus on only one experiment, that is, experiment $AB$ (similar calculations and results hold for the other experiments). Experiment $AB$ can be realised by considering two examples of {\it Animal}, namely, {\it Horse} and {\it Bear}, and two examples of {\it Acts}, namely, {\it Growls} and {\it Whinnies}. Then, the four possible outcomes of $AB$ are obtained by juxtaposing words, so that we get the four outcomes {\it The Horse Growls} ($HG$), {\it The Horse Whinnies} (HW), {\it The Bear Growls} ($BG$), and {\it The Bear Whinnies} ($BW$), which play the role of the outcomes $A_1B_1$, $A_1B_2$, $A_2B_1$, and $A_2B_2$, respectively, of experiment $AB$ in Sect. \ref{physics}. Then, each participant has to choose which one among these four possible outcomes the participant judges as a good example of {\it The Animal Acts}. We denote by $p(HG)$, $p(HW)$, $p(BG)$ and $p(BW)$ the probability that {\it The Horse Growls}, {\it The Horse Whinnies}, {\it The Bear Growls} and {\it The Bear Whinnies}, respectively, is chosen in experiment $AB$, for a given test. These probabilities can be computed as the large number limit of relative frequencies and are reported in the intermediate columns of Table \ref{animalactsentanglements} for each of the five empirical studies.

Let us now come to the quantum representation of experiment $AB$ in the combination {\it The Animal Acts}. Since $AB$ has four possible outcomes, {\it The Animal Acts} should be represented, as a whole entity, in the Hilbert space $\mathbb{C}^{4}$ of all ordered 4-tuples of complex numbers. On the other side, a representation of {\it The Animal Acts} as a combination of the sub-entities {\it Animal} and {\it Acts} requires the composed entity to be represented in the tensor product Hilbert space $\mathbb{C}^{2} \otimes \mathbb{C}^{2}$, where an isomorphism $I_{AB}$, which can be taken as the identity operator, maps an orthonormal basis of $\mathbb{C}^{4}$ into the canonical orthonormal basis $\{ (0,1)\otimes (0,1), (0,1)\otimes (1,0), (1,0)\otimes (0,1), (1,0)\otimes (1,0) \}$. Then, the abstract nature of 
the sentence `{\it The Animal Acts}', the rotational invariance of the singlet spin state and analogy with quantum physics suggest that the composite bipartite entity {\it The Animal Acts} 
must be in a state similar to the pure entangled state $p_{s}$ represented by the unit vector in Eq. (\ref{singlet}), or by the density operator $\rho_{\Psi_{s}}=\vert\Psi_{s}\rangle \langle \Psi_{s}\vert$ in the density operator formalism. However, 
due to a lack of the presence of pure symmetry in human language, the state will be not exactly $p_{s}$, 
but it is easy to determine the state unambiguously by using the probabilities by which the outcomes were obtained, and we find 
\begin{equation} \label{contextentangled}
\vert\Psi_{AB}\rangle = \sqrt{p(HG)}\vert HG\rangle + \sqrt{p(HW)} \vert HW\rangle + \sqrt{p(BG)} \vert BG\rangle + \sqrt{p(BW)} \vert BW\rangle
\end{equation}
where $\{\vert HG\rangle, \vert HW\rangle, \vert BG\rangle, \vert BW\rangle\}$ is an orthonormal basis of eigenvectors of the product self-adjoint operator which represents experiment $AB$ in $\mathbb{C}^{2} \otimes \mathbb{C}^{2}$. In the density operator formalism, the transformed state $p_{AB}$ of {\it The Animal Acts} is represented by the density operator $\rho_{AB}=\vert\Psi_{AB}\rangle \langle \Psi_{AB}\vert$, and we know from Sect. (\ref{physics}) that $S(\rho_{AB})=0$, since $\rho_{AB}$ represents a pure state.

Next, let us calculate the von Neumann entropies associated with the states of the component conceptual entities {\it Animal} and {\it Acts} by taking the partial traces of $\rho_{AB}$ with respect to {\it Acts} and {\it Animal}, respectively, which we denote by $\rho_{{\rm Animal}}$ and $\rho_{{\rm Acts}}$.
We have proved in Ref. \citet{aertsbeltran2022} that
\begin{equation}
\rho_{{\rm Animal}} = Tr_{\rm Acts} \rho_{AB}=
\footnotesize
\begin{pmatrix}
p(HG) + p(HW) & \sqrt{p(HG)p(BG)} + \sqrt{p(HW)p(BW)} \\
\sqrt{(p(BG)p(HG)}+ \sqrt{p(BW)p(HW)} & p(BG) + p(BW) 
\end{pmatrix}
\end{equation}
and 
\begin{equation}
\rho_{{\rm Acts}} = Tr_{\rm Animal} \rho_{AB}=
\footnotesize
\begin{pmatrix}
p(HG) + p(BG) & \sqrt{p(HG)p(HW)} +  \sqrt{p(BG)p(BW)} \\
\sqrt{p(HW)p(HG)} + \sqrt{p(BW)p(BG)} & p(HW) + p(BW) 
\end{pmatrix}
\end{equation}
Finally, we take the spectral decomposition of the self-adjoint operators $\rho_{{\rm Animal}}$ and $\rho_{{\rm Acts}}$ and calculate the corresponding von Neumann entropies, using Eqs. (\ref{vonnneumann}), (\ref{shannon}) and (\ref{contextentangled}) to calculate. Table \ref{animalactsentanglements} reports the results under $S$, because one easily proves that the von Neumann entropies of the reduced operators coincide. We realise at once that the von Neumann entropy of both {\it Animal} and {\it Acts} is greater than zero in all empirical studies, which is consistent with the fact that the concepts {\it Animal} and {\it Acts} are in non-pure, or density, states. We have thus obtained a result that coincides with the one illustrate in Sect. \ref{physics} on quantum physics entities: in the pure entangled state $p_{AB}$, the entropy of {\it The Animal Acts} is less than the entropy of the sub-entities {\it Animal} and {\it Acts}. In other words, the process of conceptual combination, or composition, using the terminology of physics, reduces the entropy of the component entities. We believe that what we have found has a general validity: if we consider any text, e.g., a story-telling text, then it is reasonable to assume that the text is in a pure entangled state, because of the way `meaning' is carried by the whole text
and the mechanism in human language connected with story-telling we have called `contextual updating'. Hence, this entangled state will be associated with a null von Neumann entropy. But, this entropy will be less than the von Neumann entropy of the component concepts, 
again due to this mechanism of `contextual updating'. 

This non-classical feature that the entropy decreases as a consequence of composition allows us to conclude that the 
remark we have made on physics entanglement is also valid in cognitive realms, namely, the 
most important and significant aspect of entanglement is its structure with this mechanism of contextual updating which is realised in its `state preparation'
The measurements that produce a violation of Bell's inequalities are rather there to confirm the presence of this structure. It is in the preparation stage that `the sub-entities actively collaborate to produce a pure entangled state which reduces the uncertainty of the composite entity as compared with the uncertainty of the sub-entities'. This new way of looking at entanglement also makes it possible to establish a more profound relationship between the former and meaning, as follows.

The identification of the mechanism of contextual updating makes it 
evident 
that meaning needs to be carried by entangled states. For example, in a text, even the last word has to be entangled with all the rest of the text, because a change in the meaning in the last word may result in a complete change in the meaning of the entire text. 
This is even a technique used in, say, humor, where the punch line often occurs at the end of the story, thereby creating the humorous effect. 
Let us again make explicit the mechanism by which entanglement carries meaning 
by considering the combination {\it The Animal Acts}. The concept {\it Animal} is an abstraction of all possible animals and the concept {\it Acts} is an abstraction of all possible sounds produced by animals. But, people do not construct the meaning of {\it The Animal Acts} by separately considering abstractions of animals and abstractions of acts and then combining these abstractions. On the contrary, they take directly 
abstractions of animals making a sound, and this occurs in a `coherent way' that is represented by a superposed, more precisely, entangled state. 
This is what we called the mechanism of contextual updating.

We believe to have identified a foundationally important feature of entanglement, which persists independently of the domain, physical or cognitive, where entanglement is investigated.\footnote{The analogy is even more cogent if one accepts the `conceptuality interpretation' of quantum theory, put forward by ourselves, according to which micro-physical entities are actually concepts, rather than objects \citep{aerts2009b,aerts2014,aertsetal2020}.}.  We will see in the next section that this feature is more generally present in cultural domains.

\section{An additional example from human culture \label{culture}}
We believe that the genuine quantum structures presented in the previous sections, that is, the relationship between entanglement, entropy and uncertainty, in addition to holding in physics and cognition, are also generally valid in human culture, meant as the collection of human artefacts. In that regard, we present here an example we have introduced in Ref. \citet{aertsetal2023b} which, in our opinion, can widely extend the range of our findings in Sects. \ref{physics} and \ref{cognition}.

To make the point above more clear, we believe that any human collaboration exploits the effect we have identified in Sects. \ref{physics} and \ref{cognition}, namely, the creation of entanglement correlations in any collaboration of different individuals which decreases the entropy of the collaboration as compared to the entropies carried by each individual who participates in the collaboration. Let us consider the act of hunting as an archetypical example of what we mean by `uncertainty-reducing collaboration'. Let us suppose that two (or more) individuals are involved in a hunting somewhere in the forest. They have to collaborate to achieve their common objective, i.e. hunt the prey. Each hunter is affected by a lot of uncertainty, because the behaviour of the prey is unpredictable: the pray may suddenly run away, or move towards one of the hunters, the hunters may have to lurk behind a tree, etc. This means that the hunters have to take several actions they had not originally planned, the hunting being the only whole action that had been planned. Equivalently, there is an element of intrinsic uncertainty affecting the hunting process. Moreover, each individual, taken separately from the others, has in principle a lot of freedom, that is, potentiality, which has however to be used with the purpose of making the collaborative hunt as less uncertain as possible, that is, minimising the uncertainty of the collaborative hunting. Let us apply the terminology introduced in Sects. \ref{physics} and \ref{cognition} to this specific situation. Each hunter is in a non-pure, or density, state, characterised by high uncertainty, hence by high entropy. However, the collaboration has to produce a state of the whole hunting situation that is as close as possible to a pure entangled state, in such a way that the entropy of the hunting situation becomes as low as possible. 
We can also easily identify the mechanism we have called contextual updating in the example of hunting. Skilled hunters will indeed learn from always and at any time that it is possible to link back to the whole joint enterprise of the hunt. This will certainly and surely involve feedback mechanisms that also involve signals. Probably humans developed very complex forms of such group updating because it contributed in an obvious way to the survival of a tribe. The tribe of our ancestors that became most adept at this realization of entangled states was at the front of the evolutionary hierarchy and within the arena of natural selection. It is quite possible that some of these group updating mechanisms are not yet fully known by current science, and we suspect that further study of them may also shed light on what may be taking place in the case of measurements on physically entangled states, by offering deeper knowledge of what measurement processes, and in the case of humans themselves, perceptions, actually are. 

The lesson we can learn from the example above is that `human collaboration strives at minimizing entropy, hence uncertainty, exploiting the tool of entanglement, in the sense that each collaborator actively acts to produce an overall pure entangled state'. One would agree that this is what continuously occurs in human culture. For example, one could think to sport activities where teams are involved, where the collaborative strategies taken by the individuals forming a team consist in `entangling to reduce uncertainty', to the point that this could be taken as a definition of collaboration.

We conclude this section suggesting that such an idea of entanglement as collaboration might also have a role in the solution of the `tragedy of the commons puzzle' \citep{hardin1968}. This puzzle affects several disciplines, from economics, to ecology, philosophy, and digital sciences. The tragedy of the commons can be roughly defined as a situation in which individual users who have uncontrolled access to a shared resource acts independently on the basis of their own self-interest and, contrary to the common good of all users, provoke exhaustion of the resource through their uncoordinated actions, particularly, in the case in which there is no adequate balance between users and available resources. The issues raised by the tragedy of the commons are tightly linked with deep problems of modern society, mainly sustainability, pollution, over-population, and even the hypothesised limitation of digital resources. If put in the perspective presented in this paper, an uncoordinated action as the one in the tragedy of the commons puzzle can be seen as resulting in an uncertainty-increasing product state, where a collaborative uncertainty-decreasing action would instead be the winning strategy.

\section{Non-classical nature of quantum logical connectives \label{quantumlogic}}
The main idea developed in Sects. \ref{physics}, \ref{cognition} and \ref{culture} has an unexpected impact on quantum logic. Indeed, the presence of Bell-type correlations in the singlet spin state allows one to put forward an explanation of why some quantum logical connectives, in particular, the `inclusive and exclusive disjunctions' and the `bi-conditional', behave non-classically, even for `compatible propositions'. To this end, we briefly review in the following some aspects of quantum logic that are relevant to the purpose above.

Let $a$ and $b$ be two propositions of quantum logic and let us denote by $\neg$, $\land$, $\lor$, $\veebar$, $\rightarrow$, and $\leftrightarrow$, the connectives of quantum logic corresponding to the connectives of `negation', `conjunction', `inclusive negation', `exclusive negation', `conditional', and `bi-conditional', respectively.\footnote{We stress that, despite they are denoted by the same symbols as in classical logic, quantum logical connectives formalise empirical relations between propositions, hence they are fundamentally different from the corresponding classical logical connectives.} We have learned from quantum logic that a proposition $a$ is represented by the orthogonal projection operator $P_a$ on the Hilbert space $\mathscr H$ associated with the entity $S$ which $a$ refers to. We then say that `$a$ is true' if the state $p$ is such that, whenever $a$ undergoes a `yes-no measurement' $\alpha$, the outcome `yes' can be predicted with certainty, that is, with probability equal to 1. Coming to the Hilbert space representation, if the state $p$ is represented by the unit vector $\vert \psi_p\rangle$, the proposition $a$ is true if and only if $P_a\vert \psi_p\rangle=\vert \psi_p\rangle$. In addition, we know that the quantum logical connectives $\land$ and $\rightarrow$ behave classically, while the connective $\neg$ behaves non-classically, in the sense that it may happen that neither $a$ nor $\neg a$ are true \citep{bc1981,dcgg2004}.

We intend to provide an explanation of why the quantum logical connectives $\lor$, $\veebar$ and $\leftrightarrow$ have an especially non-classical nature. Our analysis relies on the investigation in Refs. \citet{aerts1982} and \citet{aertsdhondtgabora2000} but adds to it the crucial ingredient of the new role played by entanglement presented in this paper. Let us now suppose that the propositions $a$ and $b$, represented by the orthogonal projection operators $P_a$ and $P_b$, respectively, refer to a composite entity $S$ made up of the sub-entities $S_1$ and $S_2$, and also suppose are $a$ and $b$ compatible, that is, $P_aP_b=P_bP_a$. Let $\alpha$ and $\beta$ be the yes-no measurements testing $a$ and $b$, respectively. Compatibility of $a$ and $b$ means that $\alpha$ and $\beta$ can be performed together. Let $\alpha \land \beta$ be the measurement testing $a$ and $b$ together: $\alpha \land \beta$ has four outcomes which correspond to the fact that both $\alpha$ and $\beta$ can give `yes' or `no' as outcomes. Considering the measurement $\alpha \land \beta$, we then say that:

(i) the conjunction $a \land b$ is true if and only if the state $p$ of $S$ is such that in the measurement of $\alpha \land \beta$ we obtain with certainty `yes' for $\alpha$ and `yes' for $\beta$;

(ii) the inclusive disjunction $a \lor b$ is true if and only if the state $p$ of $S$ is such that in the measurement of $\alpha \land \beta$ we obtain with certainty `yes' for $\alpha$ and `yes' for $\beta$, or `yes' for $\alpha$ and `no' for $\beta$, or `no' for $\alpha$ and `yes' for $\beta$;

(iii) the exclusive disjunction $a \veebar b$ is true if and only if the state $p$ of $S$ is such that in the measurement of $\alpha \land \beta$ we obtain with certainty `yes' for $\alpha$ and `no' for $\beta$, or `no' for $\alpha$ and `yes' for $\beta$;

(iv) the exclusive disjunction $a \leftrightarrow b$ is true if and only if the state $p$ of $S$ is such that in the measurement of $\alpha \land \beta$ we obtain with certainty `yes' for $\alpha$ and `yes' for $\beta$, or `no' for $\alpha$ and `no' for $\beta$.

One would intuitively think, by analogy with the truth tables of the classical logical connectives corresponding to $\land$, $\lor$, $\veebar$, and $\leftrightarrow$, that:

(1) $a \land b$ is true if and only if $a$ is true and $b$ is true;

(2) $a \lor b$ is true if and only if $a$ is true and $b$ is true, or $a$ is true and $b$ is false, or $a$ is false and $b$ is true;

(3) $a \veebar b$ is true if and only if $a$ is true and $b$ is false, or $a$ is false and $b$ is true;

(4) $a \leftrightarrow b$ is true if and only if $a$ is true and $b$ is true, or $a$ is false and $b$ is false.

This is not the case if $S$ is the composite entity made up of two spin-1/2 quantum entities prepared in the singlet spin state $p_{s}$, $a$ is the proposition corresponding to the measurement of spin along an arbitrary direction, e.g., direction $A$, Sect. \ref{physics}, and $b$ is the proposition corresponding to the measurement of spin along the opposite direction, e.g., $-A$. In this case, indeed, the `perfect Bell-type anti-correlation' in the singlet spin state 
implies that the measurement $\alpha \land \beta$ `always' leads to the outcome `yes' for $\alpha$ and `no' for $\beta$, or `no' for $\alpha$ and `yes' for $\beta$. As a consequence, $a \lor b$ and $a \veebar b$ are true, while $a \land b$ and $a \leftrightarrow b$ are false. But, neither $a$ nor $b$ are either true or false: they are completely indeterminate in the state $p_s$. This entails that, while (1) holds, (2), (3) and (4) do not hold.

It follows from the analysis above that it is exactly the presence of Bell-type correlations which makes the quantum logical connectives corresponding to inclusive and exclusive disjunctions and the bi-conditional behave in a non-classical way. This can be naturally explained if one accepts that entanglement is a process of reduction of uncertainty. This reduction of uncertainty makes it possible the composite proposition, be it $a \lor b$, $a \veebar b$ or $a \leftrightarrow b$, have a definite truth value without neither of the component propositions $a$ and $b$ having a truth value. Hence, entanglement is responsible of a `more classical logical structure of propositions'. That the component propositions have no truth value is a consequence of the fact that, in the singlet spin state of the composite entity $S$, the sub-entities $S_1$ and $S_2$ do not possess any property, hence they lose, in some sense, their identity as a result of the composition.

It is finally important to stress that, though this result has been derived in the case of micro-physical entities and realm, it equally holds in the cognitive and, more generally, cultural realms. This demonstrates that entanglement has a profound impact on the logical structure of entities, independently of their origin.

\end{document}